\documentclass{article}

\usepackage{arxiv}

\usepackage[utf8]{inputenc} 
\usepackage[T1]{fontenc}    
\usepackage{hyperref}       
\usepackage{url}            
\usepackage{booktabs}       
\usepackage{amsfonts}       
\usepackage{nicefrac}       
\usepackage{microtype}      
\usepackage{lipsum}
\usepackage{graphicx}

\title{Intrinsic PUF Instance on Non-Volatile NAND Flash Memory}

\author{
  Surbhi~Vasudeva \\
  Department of Computer Engineering\\
  San Jose State University\\
  San Jose, CA 95192 \\
  \texttt{surbhi.vasudeva.edu} \\
    \And
  Sara~Tehranipoor\\
   Department of Electrical and Computer Engineering\\
  Santa Clara University\\
  Santa Clara, CA 95050 \\
  \texttt{ftehranipoor@scu.edu} \\
  \And
  Nima~Karimian\\
  Department of Computer Engineering\\
  San Jose State University\\
  San Jose, CA 95192 \\
  \texttt{nima.karimian@sjsu.edu} \\
}

\begin{document}
\maketitle

\begin{abstract}
	
Embedded systems or micro controller based modules have become increasingly prevalent in our daily lives. However, the security of embedded devices as well as the authenticity of hardware has become an increasing concern within the growing Internet of Things (IoT) space. In this paper we setup an experiment environment where SLC flash program disturbance is observed.We discovered that intra-page disturbance is easier to be produced than inter-page disturbance. We also observed that adjacent pages are paired in (2n, 2n+1) manner, and disturbance only occurs within a pair. Lastly, we found that as page number increases from 0 to 63, it becomes more difficult to observe the first bit flip within a page, and thus more difficult to achieve the disturbance stable state.
  
\end{abstract}

\section{Introduction and Motivation}
Technology is prevalent in every field today. Lives have become so much dependent on electronic devices. The electronic devices in use are mostly part of a bigger network - the Internet of Things (IoT)~\cite{hassan2017internett, tehranipoor2018towards}, in the form of sensors or nodes. The information reaches its final destination for analysis or application through the cloud. Within this setup, there is no guarantee of the integrity of the data. Authentication~\cite{tehranipoor2017exploringg, yan2015novelll, karimian2019dramnett} appears to be the solution to this problem. In the past, when it became difficult to distinguish/identify humans based on their physical traits, biometric authentication~\cite{karimian2016evolving, karimian2019heartid} was invented. Similarly, electronic devices, for example, RFIDs, look alike. The only difference is the ID stored in the chip memory, which is vulnerable and prone to attack. Also, it can be modified easily~\cite{wortman2020exploring}. As an example, For costlier hardware devices that are produced in bulk, there are so many people involved in the production chain. At any point in time, people can decide to use the product for their benefit and introduce the product in the market while solely enjoying the profits without sharing with the parent company. Such an act is unacceptable for a reputed company. The company’s reputation and millions of dollars could be at stake. What if the device could have a fingerprint just like human beings? Then it becomes easier to identify the device uniquely without any ID embedded in it. How this is possible is due to the inherent nature of silicon devices as the physical variations that happen on physical devices cannot be reproduced through manufacturing. Hardware security primitives are potential solutions~\cite{yan2020bittt, wortman2018p2mmm, tehranipoor2018dvftt, eckert2017drnggg, tehranipoor2016robust, tehranipoor2017study, tehranipoor2017designnn, lyp2021lish} to secure consumer electronic devices~\cite{tehranipoor2018loww}. Physically unclonable function (PUF) is the technique of digital fingerprint on physical semiconductor devices. This fingerprint remains almost unaffected due to any temperature, humidity, or stability variations~\cite{aguirre2020systematiccc, anagnostopoulos2018addressingg, tehranipoor2017investigation}.
There has been research that investigates PUF from volatile memories which can be found in~\cite{tehranipoor2016dram,yue2020drammm, Holcomb2008, karimian2019generateee, anagnostopoulos2018securing, anagnostopoulos2018overvieww, anagnostopoulos2017insights, tehranipoor2015dram} and non-volatile memories in~\cite{gordon2021flashsh}.
\par
Most electronic devices need memory to store any information. From the list of all the available memories, NAND Flash is the widely used one where many different applications such as USB drives, media players, digital cameras, and smartphones use it. That is because it has lower power consumption, has shorter write and erase time (high access speed), and has a low bit cost. With the introduction of PUF, IoT security especially in healthcare domain~\cite{wortman2017proposing, tehranipoor2017investigation}, can be enhanced as the PUF provides a unique ID, which is unclonable.
Secondly, if data is stored with PUF encrypted on memory, the memory is secured as there is no chance of decryption without the PUF. Third, PUF behaves like a random function, i.e., generating random output values, and there are numerous applications of a good and reliable random number generator. This function is unpredictable for even an attacker with physical access to the system. Also, It is impossible to produce a copy of the same physical system even when the functionality is known. Physical Unclonable Functions (PUFs) provide a unique signature~\cite{yan2017phasee, yan2016pufff}. Where the signature is the bits that have been stable no matter how many read and write operations applied to that cell. An extracted signature can be used to authenticate a chip and produce a random number that can be used in cryptography keys. The variation takes place between blocks and between pages. That means the chip can have multiple unique signatures in different locations, and different methods can extract them. Some PUF models are vulnerable to attacks~\cite{pugazhenthi2019dlaa}.
\par
In this paper, we implemented PUF extraction using the Samsung NAND Flash Memory (K9F1G08U0E) with the FMC interface of the experimental discovery board (STM32F429ZIT6). The software used is STMCUBE32IDE with C programming language. We researched different program disturb algorithms and implemented them on  NAND flash to extract the PUF and observe the performance of the same.

Specifically, we have the following contributions in this paper:
\begin{itemize}
   \item Observed SLC Single Page Program Disturbance.
   \item Observed SLC Multi-Page Program Disturbance.
   \item Observed SLC pages within a block are paired.

\end{itemize}

The remainder of the paper is organized as follows:  Section II are describing related work, reviews the basics of NAND FLASH, followed by a description of PUF and various techniques to identify PUF in Section III. Section  IV  describes the methodology and approach followed to obtain PUF. Section V demonstrates the hardware and software setup, along with operations performed.  Section  VI shows the results of our  NAND Flash PUF implementations, and finally, we close with conclusions in Section VII.

\section{Related Works}
Shijie et al.~\cite{jia2015extracting} introduced a PUF-based key generator for NAND flash chips. They proposed three methods to extract raw PUF output numbers including position map and partial programming and partial erasure to select reliable keys. Prabhu et al.~\cite{prabhu2011extracting} proposed seven NAND flash PUFs based on program disturb, read disturb and program operation latency. They have experimented with fourteen devices. They have used Pearson correlation to evaluate and measure the robustness of the signatures. Their results showed NAND flash based on program latency and program disturb are the most useful ones. They have observed that NFPUF based on program
disturb is the best one to distinguish between different chips but takes more time for extraction and NFPUF based on program latency is the fastest among all PUFs. Cai et al.~\cite{cai2017vulnerabilities} demonstrated the effects of two-step programming on MLC NAND flash. Their work exposed the dangers of two-step programming and how it could be exploited. The process of two-step programming involves programming a single cell’s LSB and MSB at different times. Most of the existing work is limited to analyzing and generating PUF for a single NAND flash chip either SLC or MLC. In this paper, we will observe SLC Single Page Program disturbance and SLC pages within a block are paired.

\section{Background}
\subsection{NAND Flash}
The NAND Flash cell is made of a floating gate transistor, where the electrons held by the floating gate decide the threshold voltage ($V_{th}$) of the cell. Normally to switch on an NMOS transistor we need $V_g > V_{th}$. During the program stage, if we inject electrons in the floating gate, we will require a relatively higher voltage ($V_g$) to offset the negative charge of electrons ( and turn on the channel ($V_{g}-V_{\delta}>V_{t h}$). If there are no electrons captured in the floating gate a relatively small gate voltage is sufficient to make $V_g > V_{th}$ and turn on the transistor.
NAND Flash is non-volatile and can retain data even when power is not supplied. When power is detached from NAND flash memory, a metal-oxide-semiconductor provides an extra charge to the memory cell that helps to keep the data. This metal-oxide-semiconductor is known as the Floating gate. Regular kinds of NAND flash storage incorporate SLC, MLC, TLC, QLC, and 3D NAND. What isolates each type is the number of bits every cell employs. The more bits put away in every cell, the more affordable the NAND flash would cost. NAND Flash stores data in blocks. The basic unit of erase operation is this block, whereas the basic unit of a write operation is a page. Each page consists of a data area and a spare area. This data area and spare area vary, and it depends upon flash size and manufacturer. This spare area is mainly used to store the error correction algorithm or to store the map table-related information in a page layout. Basic operations performed in NAND Flash are to read or program a page or erase a block.

\begin{figure*}
\caption{Flash Memory Organization}
\includegraphics[width=0.85\linewidth]{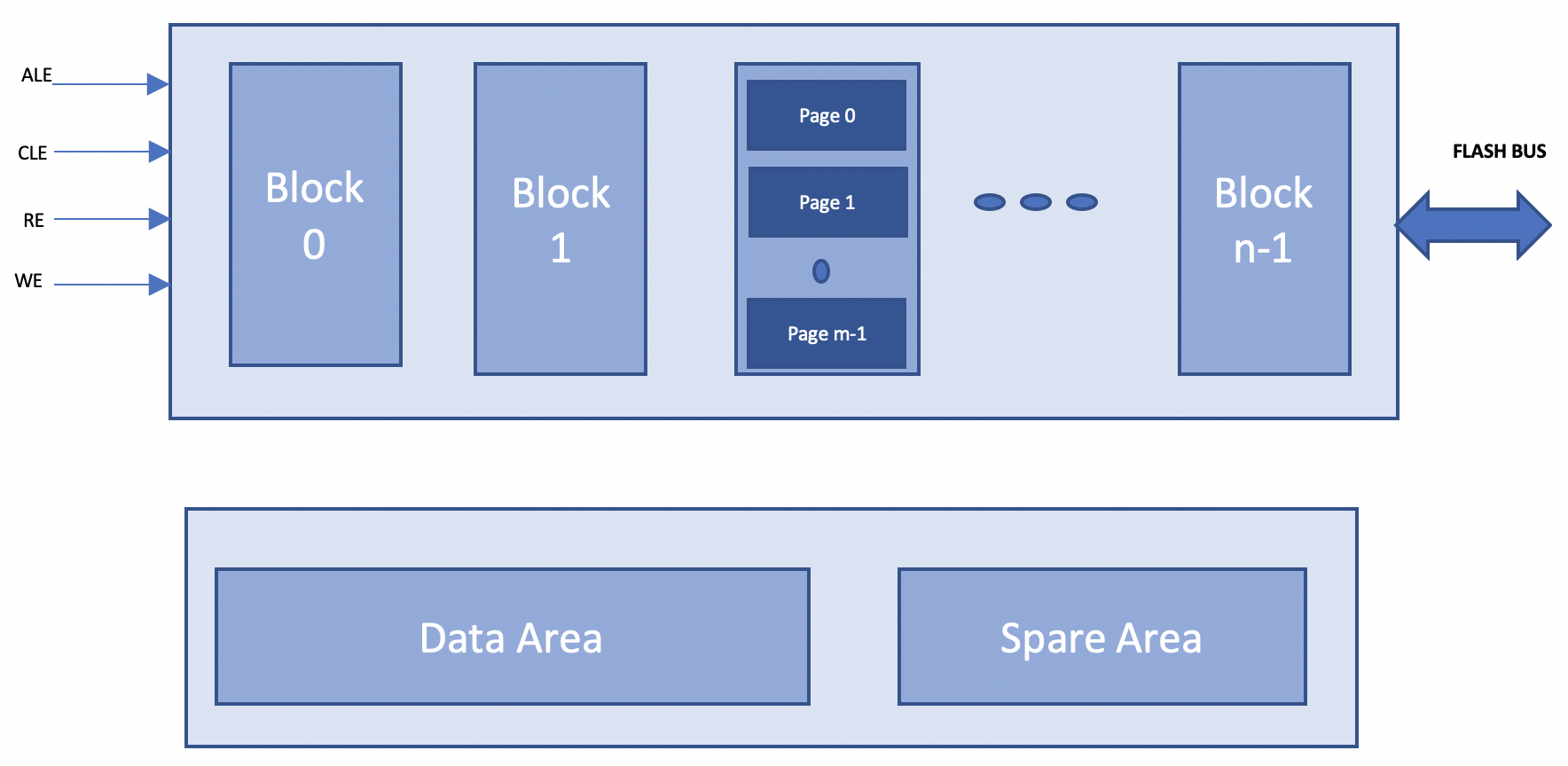}
\end{figure*}

\subsection{PUF within NAND Flash Memories}
Flash cells are densely packed, so process variations in terms of feature geometry can have a significant effect on the coupling between nearby cells, or the same cell behavior. Because of these variations, some cells might be more or less susceptible to write/read disturb/wear. Physical Unclonable Functions (PUFs) provide a unique signature. Where the signature is the bits that have been stable no matter how many read and write operations applied to that cell. An extracted signature can be used to authenticate a chip and produce a random number that can be used in cryptography keys. The variation takes place between blocks and between pages. That means the chip can have multiple unique signatures in different locations, and different methods can extract them. Physical Unclonable Function helps to prevent the storage of secret keys in the storage memory.
\subsection{PUF Generation Techniques}
There are different methods to extract raw PUF output numbers from NAND Flash memory chips, which are Program disturb, Read disturb, and Program operation latency. In Program disturb, a block will be erased and then repeatedly program one page in the block. Between each Program, the adjacent pages should be read for Program disturb-induced errors. For each bit on the adjacent page, we have recorded the number of programs required to cause the first-bit error. Those values comprise the signature. Extracting a signature requires one erase operation and many program operations, and these programs don’t affect reliability. The disadvantage of program disturb is that it causes irreparable damage to the page, and it is quite slow. \par
The second method is to read disturb. Within this technique, we first erase the entire block, then program it with random data. Next, we read from each page several million times to induce a read-disturb. After every 1000 read iterations, we checked each page in the block for errors. If we observed an error on a page, we recorded the bit, the page, and the cycle of the error. This process is repeated 10 million times. Thus, we use the read cycle counts for all the bits in the block as the signature. The Read disturbs are less destructive, and it causes less noise disturbance in signature than the Program disturbance technique, yet it is slow. \par
Program operation latency is another method that is used to extract signatures. Within this technique, we program one bit at a time on a page and record the latency for each operation. There are two types of operation latency we can introduce, which are a single page and multiple pages. The advantage of this method is that it is fast, and it dost not cause wearing the Flash device. On the other hand, the information content of the signature is low and is easier to forge.

\section{Methodology and Approach}
\subsection{Algorithm}
After looking into different methods to extract the signature. We decided to use the program disturb in our project. We used this method because we can notice the distribution happening more frequently in the cells than using read disturb. That is because program disturb can appear in earlier iterations, and it needs approximately ten thousand iterations to be able to see it. Also, it could appear earlier than that. On the other hand, with read disturb, we can not see a lot of distribution happening within ten thousand times. it needs to be run for longer iterations to observe the read disturb.
The general idea to achieve the program disturbance is by first erasing the targeted block. Then, choose a page within that block and program it repeatedly for approximately ten thousand cycles. In each cycle, we will read the adjacent page for any errors. If there is any error i.e a bit is flipped. We will record the cycle number corresponding to that bit in our signature. We will continue to monitor for other bits and then for the number of cycles shows the pseudo-code that is used to achieve this approach.

\begin{figure*}
\includegraphics[width=0.7\linewidth]{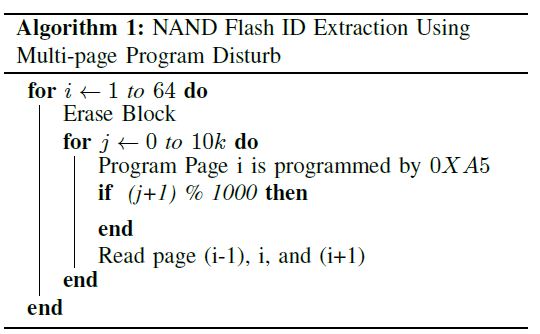}
\end{figure*}

\begin{figure*}
\includegraphics[width=0.7\linewidth]{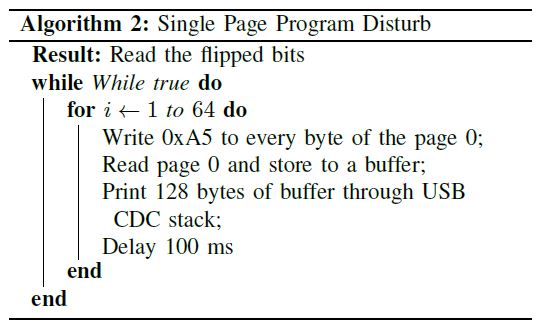}
\end{figure*}

\subsection{Procedure}
We found stable bits in the NAND Flash memory in different ways, on the same page, on the adjacent page, and on all pages within the block. To start with, the first approach, which is finding the stable bits on the same page. We erased block zero and program page one with value AA in hexadecimal. After that, we monitored changes in bits that were occurring on the same page. We did that by comparing the values of the bits on the same page with AA. If the value changes, that means that the bit is not stable. Otherwise, they are stable bits, and we stored them in an array. After that, we repeated the program operation and the steps after it for ten thousand iterations. Then, by the end of the ten thousand iterations, we had stable bits that generate a PUF. \par
The second approach is similar to the first one, but this time we are reading the adjacent pages. In this, we erased block zero, and then programmed page two with value AA in hexadecimal. After that, we read the adjacent pages, which are page one and page three. In these adjacent pages, we compared the bits inside them with the value of FF (FF is the value generated due to the erase operation; it set all bits to ones). If it is the same value, that means it is stable. If it changed, that means it is not stable. After that, we stored the stable bits in an array and keep comparing them at each iteration ten thousand times. Where in each iteration, we write to the same page, which is page 2, and read bits from page 0 and page 3.\par
 In the last approach, erased block zero and programmed all pages with AA. Then choose page one and program it again with AA ten thousand times. Then, read the adjacent pages, which are page zero and page two—compare the bits on them with the value of AA. If it changed, it is not stable; otherwise, it is the stable bit and stores them in an array. After that, I moved to the next page, which is page two, and programs it with AA ten thousand times. Next, read the adjacent page of it, which is page one and page three, and compare the value it has with AA. If it did not change, we would put them in an array. Applied the same steps to all pages in the block, and there are 64 pages. ECC needs to be disabled. In case ECC is enabled, then it will be corrected automatically.

\section{Hardware Setup and Implementation}
\subsection{Hardware Connection}
The hardware connections between the STM32F429I and the NAND Flash are made using female jumper wires. Connections to a total of 16 pins on NAND Flash is required. Among the 16, there are 8 data pins, 6 control pins, Vcc, and GND. The same 8 data pins on the NAND Flash would be used to pass command, address, and data. The 6 control pins on the NAND Flash are CLE(Command Latch Enable), ALE(Address Latch Enable), R/B(Ready/Busy Output), WE(Write Enable), RE(Read Enable), CE(Chip Enable). Power is supplied to NAND Flash through a microcontroller by connecting its Vcc 3V to the Vcc of the NAND Flash and GND of the microcontroller to GND of NAND Flash. Corresponding FMC(Flexible Memory Controller) pins from the microcontroller are connected to NAND Flash control and data pins. STM32CubeIDE is used to develop the software code in the C programming language. STM32CubeIDE has the STM32CubeMX graphic tool to visualize the board, select pins and configure them as well. The STM32CubeIDE provides the facility to generate definitions and library files with the configured pins from the graphic tool. 

\begin{figure}
\center
\includegraphics[width=0.85\linewidth]{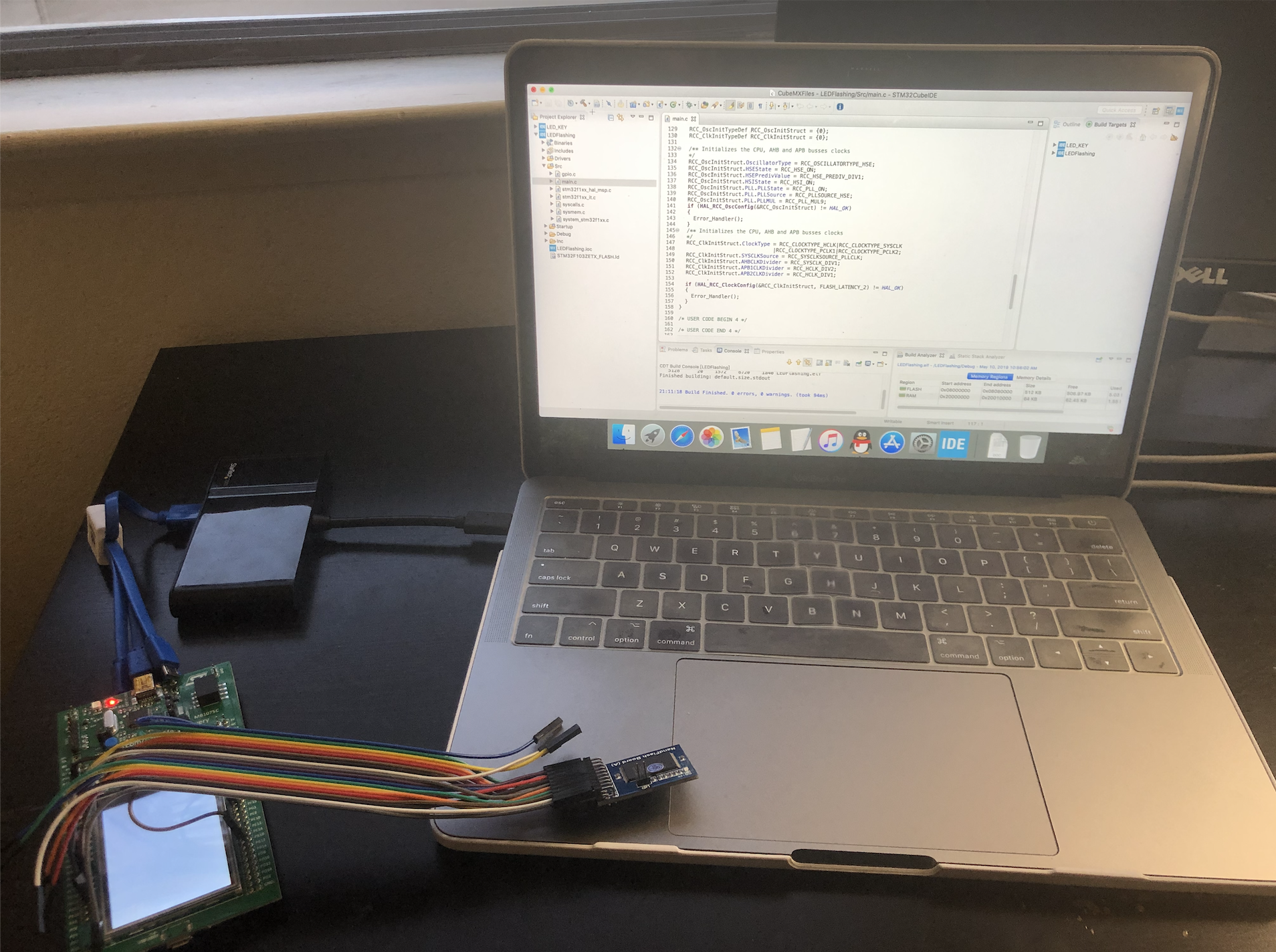}
\caption{STM32Board connection with machine and flash memory.}
\label{STM32Board1-fig}
\end{figure}

\subsection{Software Setup}
STM32CubeIDE is used to develop the software code in the C programming language. STM32CubeIDE has the STM32CubeMX graphic tool to visualize the board, select pins and configure them as well. The STM32CubeIDE provides the facility to generate definitions and library files with the configured pins from the graphic tool. 

\subsection{Configuration}
As part of the configuration, first, the Hardware Abstraction Layer (HAL) has been initialized, followed by System clock configuration. Further other peripherals like GPIO, FMC, and USART1 are initialized. A NAND Controller instance is created, and the required setup values are configured. These include SetupTime, WaitTime, NandBank, ECCcomputation, BlockSize, PageSize, and others.

\subsection{ERASE and READ Operations}
Erase operation is a must before programming the page. Erase Operation takes place at block level. Erase operation is initiated by issuing Erase setup command 60h, followed by row address. The Erase confirm command D0h initiates internal Block Erase operation. R/B pin is set to low. The status is checked by issuing command 70h. Value 0 in Write Status bit (I/O 0) indicates that Erase is completed whereas value 1 indicates an error in Erase operation. A Read operation is initiated by writing the opcode 0x00. The CLE (command latch enable) pin must be asserted HIGH while simultaneously asserting the CE (chip enable) pin LOW. This command alerts the flash device to prepare the data buffer registers to read from an incoming address. The row address and column address is then sent after the 0x00 opcode and the address is written on consecutive rising edges of the WE (write enable) signals. It requires 4 bytes to locate which page to read from. 2 bytes for the row address and 2 bytes column address. This yields a total of 64k addressable row pages and 64k addressable columns. However the last 4 bits of the column address are not used because there are a total 2112 bytes in the column space (2096 bytes per page + 64 bytes of spare area for ECC). This means that 12 bits are required to address the columns. The completion of writing the address is indicated by sending the opcode 0x30. Now the device now knows which bytes to read from its flash array. The data from the address is loaded onto the page register and sent through each of the 8 I/O pins to be read by the STM32F4 micro-controller. Each byte of data is sent on consecutive falling edges of the RE(read enable) signal.
\subsection{PROGRAM Operation}

\subsection{PROGRAM Operation}
NAND Flash is programmed at page level. Program operation is initiated by issuing Serial data input command 80h, followed by column and row addresses. Then, Transmitter buffer is set and is passed in the next cycle. The transmitter buffer consists of the data to be written on the page.   The NAND instance is initialized with frequency, setup timing, hold timing, etc. The address consists of a column address and row address. Read operation is performed to verify the data written to the NAND Flash. A command input of value 00h is passed to perform read page operation. Column address and row address are passed in the next 4 cycles. Data is read in the next cycle. Read Page output is in the receiver buffer. The NAND instance is initialized with frequency, setup timing, hold timing, etc.  The receiver buffer holds the data read from the page. This initiates programming process. R/B pin is set to low. The status of write operation is checked by issuing command 70h. Value 0 in Write status bit (I/O 0) indicates that a program is completed whereas value 1 indicates an error in program operation. A command input of value 80h is passed, which determines it is a write operation. Column address and row address are passed in the next 4 cycles. 
\par
1) PUF observing the same page: Program disturb technique on the same page is observed by programming and simultaneously reading the same page for 10,000 iterations. Bit flips are observed after 4000 iterations. The continuous programming to the same page causes disturbs, which can be seen in reading operation after around 4000 iterations.  After block erase operation, the transmitter buffer is loaded with 0XAA for page size, which is 2048 bytes. The program and read operations are performed in a loop for 10,000 iterations. The program operations are invoked with the NAND controller handle, address setup for page 2, transmitter buffer loaded with 0XAA, and number of pages to write as 1. The read operation is invoked with the same NAND controller handle, address, receiver buffer, and the number of pages to read as 1. The data in the receiver buffer is then printed to the serial output using the printf statement. At the end of the for loop, the receiver buffer holds the data read from page 2 in last iteration. The receiver buffer data is then compared to 0XAA which is the value programmed. The indexes of the bytes whose value matches to 0XAA are then identified to be stable bytes.
\par
2) PUF observing the adjacent pages: As per our research, we found out that we can obtain PUF within flash memory using page disturb technique for adjacent pages. The initial algorithm we decided on for this technique is shown as follows: As per this algorithm, we identified block number 0 to be our target block and page number 3 to be our target page. Following this, we identified page number 2 and page number 4 to be the two adjacent pages to observe any bit flips. First of all, the target block 0 was erased, then we write pages number 2, 3, and 4 with the same data (0xAA) once. Page number 3 is written to for 10,000 times and observed for bit flips by comparing the RxBuffer data byte with the data initially written to it. If the data within the byte was equal to 0xAA, we marked that bit as stable and noted the cycle number for that particular bit location in our stableBits array. Finally, the stableBits array was populated after checking for every byte for any flips. In this algorithm, the adjacent page read for bit flips is not verified bit-wise. With one bit flip in a byte, the whole byte is disregarded. Hence, individual stable bits are not being captured. As a solution to this, We utilized the XOR property - if two variables having the same value are XORed together, the output will be false. The output will be true only if the two variables are not the same. The second logical property we utilized is masking provided by AND operation. The change in a bit is observed by masking each bit. The location property of a bit within a byte is used for each iteration. After running this code, we were able to obtain the number of stable bits based on the iteration number where the other bits were flipped.
\par
3) PUF observing a number of pages:  In this part of the experiment, the search for stable bits was extended to all the 64 pages of the block. To create a program disturb, the number of write iterations was set to 5000.
The new challenge when dealing with multiple pages was data gathering. Coming up with a unique fingerprint requires complex processing. Doing all this calculation within the limits of the Discovery board was not only challenging but also time-consuming. Thus, the data was gathered in a raw format. This raw data was later processed using a python script to come up with the PUF instance.
The code snippet shows the algorithm to extract the stable bits. The post-processing of the gathered data involved the following steps: 

\begin{enumerate}
\item {Identify stable bits: For each page in the block, the stable bits were gathered across multiple passes.}
\item{Find Correlation: For each page in the block, the stable bits were gathered across multiple passes.} 
\item{Generate PUF Instance with the highest co-relation: For each page in the block, the stable bits were gathered across multiple passes.}
\end{enumerate}

\section{Results}
\subsection{PUF Identification}
1. PUF observing the same page: 
Algorithm 1 is programmed into the discovery board and will demonstrate single page program disturb. The same page program disturbs technique experimented on two different NAND Flash devices. The number of iterations and the page programmed are also kept the same for both the devices. The number of iterations experimented on both devices is 10,000. The page programmed and observed for program disturb is page 2. The stable indexes bytes identified in both the NAND Flash devices are shown in Figure. The number of stable bytes identified in each device is different, and also the indexes of the stable bytes varied. From the experiment, we found after about 306 iterations the bits start to flip and become much worse afterward. The explanation to the above observation is as following:
Flash can only be write from 1 to 0, once it is 0, it can only be toggled to 1 by erasing. Initially each flash cell is programmed as 1s, and then 0xA5 is repeatedly written to each byte. Ideally, the read output should be 0xA5 consistently. But after some iteration, some bits flipped from 1 to 0 and can never get back, as we are not involving erase operation for each iteration. Consequently, the flipped bits will accumulating to a more and more chaotic status.

\par
2. PUF observing the adjacent page: 
After performing 10,000 writes and reads, we were able to find the number of stable Bytes. We performed a similar operation on two devices to observe if the output obtained is unique or not. We observed that for both the devices, all the bytes on page number 2 were stable, whereas a limited number of stable were found on page number 4. The image shows the number of stable bytes on page number 4, where continuous writes were performed on page number 3.\par 

\begin{figure}
\centering
\includegraphics[width=0.8\linewidth]{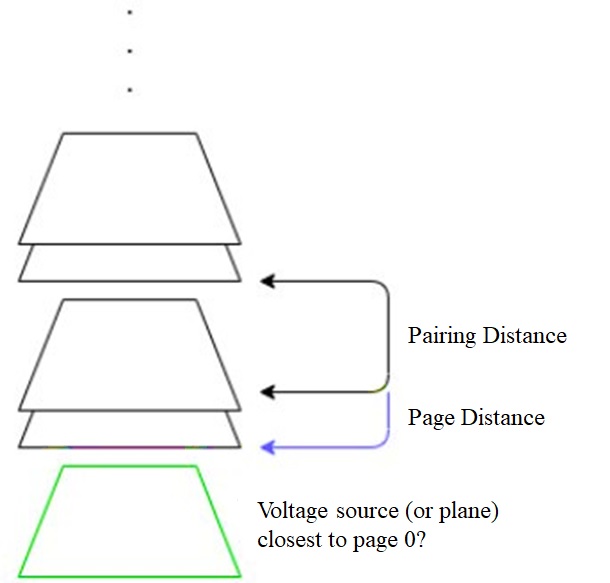}
\caption{Speculated Physical Layer of Block on Samsung SLC.}
\label{samsung}
\end{figure}

\begin{figure}
\center
\includegraphics[width=0.8\linewidth]{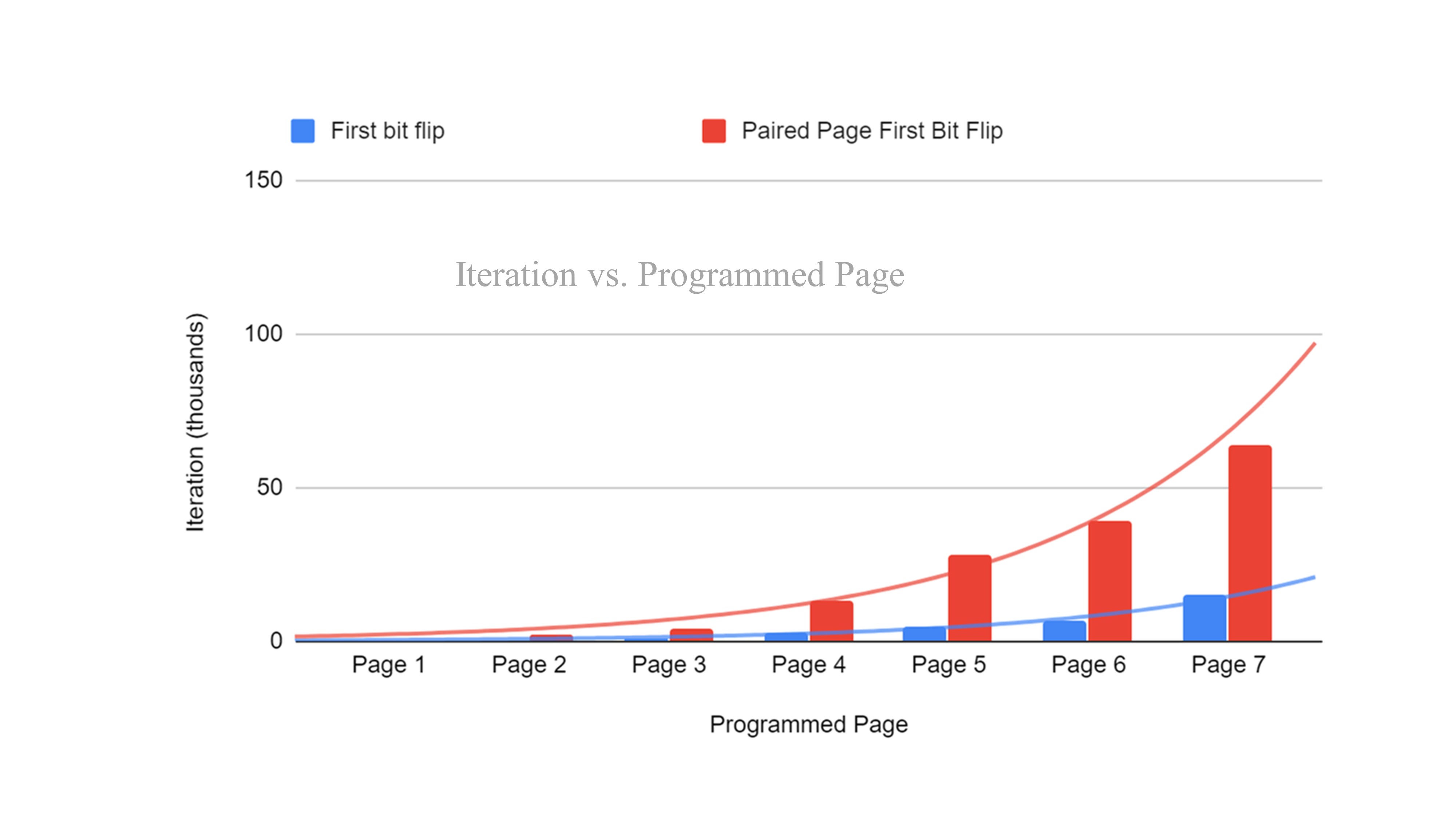}
\caption{Estimating Bit Flip Pattern.}
\label{Estimating}
\end{figure}
\begin{figure}
\center
\includegraphics[width=.8\linewidth]{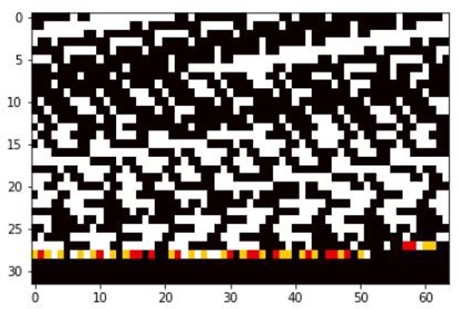}
\caption{Page 1 signature after 100k multi-page program writes.}
\label{Page1}
\end{figure}
\begin{figure}
\center
\includegraphics[width=0.8\linewidth]{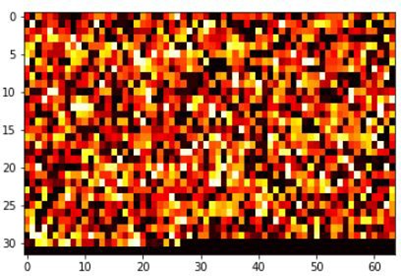}
\caption{Page 2 signature after 100k multi-page program writes.}
\label{Page2}
\end{figure}
3. PUF observing several pages:
Algorithm 2 is used to observe intra-page and adjacent-page disturbance on pages 1 through 64. However, in this study, we only evaluated 7 pages. Table I displays the iteration at which Page 0 gets affected from Page 1 which is less than 1,000 write iterations. However, Page 2 is undisturbed which leads to the conclusion that only pages that undergo disturbance happen in pairs. This is dependent on how Samsung's SLC NAND flash memory is layered physically and the spacing between two out of three pages are within reachable distance for multi page disturbance seen in Fig.~\ref{samsung} where there is a page and pairing distance. The page distance is close enough such that bit flipping occurs while pairing distance is far enough that one pair does not disturb the other. Pages become more resilient to bit-flipping as they stray away from page 0. Without knowing the physical layer of how Samsung built it’s SLC, we speculate that page 0 is the closest to the voltage source (or plane) of the block since we know that pairs are stacked vertically. The driving factor for how fast a program disturbance occurs aside from how physically close the pages are is from heat (of this voltage source). As we climb the pages vertically, heat dissipates thus causing program disturbances to increase exponentially. This causes the extraction of signatures to become much more tedious in following manuscript and will be discussed as we compare signature extraction of pages 1 and 2.
After recording data into log files, we develop a Python script to analyze and plot our observations. Our analysis proved that the pages are physically programmed in pairs which follows the equation (2n, 2n + 1), where n is the pair number. Programming to page 0 only causes disturbances on page 1, programming on page 2 only affects page 3 and so on. Further analysis shows that observations for program disturbs require more iterations as the pages increase from 0. This is seen from plotting Table I in Fig.~\ref{Estimating}. An exponential curve fit is applied to both inter- and intra- page disturbances which allows for estimating when a page and it’s pair will encounter first bit flips.

The black colored entries in the heatmap show bytes that have completely flipped. Meaning after erasing the page, the all bits in a byte have changed from 0xFF to 0x00. The white entries show erased bytes that remained 0xFF after page erase. Using the non black colored entries, we can use these markings to show the bytes and bits of a page that have not flipped from program disturbs. After 83k iterations, the signature became stable. The colored entries represent bit flips within a byte. In Page 2’s signature it is quite evident that all the bits in each of the 2048 bytes do not completely flip. Only a few of the 8 bits flip. We can use the colored entries to show the bytes. Similar to page 1’s heatmap where the black colored entries represent the flipped bytes and white entries show stable bytes, we observe that 100k iterations is not enough program writes to observe a stable signature specifically for page 2. Examples can be seen in Fig.~\ref{Page1} and Fig.~\ref{Page2}. The higher numbers are represented by warmer colors, which are indicative of bits that were highly resistant to disturbance and only flipped after running the experiment for high number of cycles.

\begin{table*}[]
\small
\centering
\begin{tabular}{|l|c|c|c|c|c|c|c|}
\hline
Program on Page n & Page 1 & Page 2 & Page 3 & Page 4 & Page 5 & Page 6 & Page 7 \\ \hline
Page (n-1)/k & \textless{}1 & No bit flip & 4 & No bit flip & 28 & No bit flip & 64 \\ \hline
Page n/k & \textless{}1 & \textless{}1 & 1 & 3 & 5 & 7 & 15 \\ \hline
Page (n+1)/k & No bit flip & 2 & No bit flip & 13 & No bit flip & 39 & No bit flip \\ \hline
\end{tabular}
\caption{First bit flip for disturbed pages 1 to 7. Column 1 is showing number of iteration for the First bit flip.}
\end{table*}

\section{Conclusion and Future Work}
In conclusion, this project helps understand the PUF and its implementation. The program disturb technique is used to extract PUF. As Read disturb technique needs more than a million reads to observe bit flips, program disturb is performed. Read or Program Disturb on SLC NAND flash memory causes lesser bit flips as compared to MLC NAND flash memory device. In this project, unique signatures in different NAND flash memory are extracted. PUF is captured on the same page, in the adjacent pages, and all pages. The result is compared with two different NAND flash memory devices to compare the signatures. PUF is identified to be unique for each NAND flash memory. Thus, PUF extracted, could be used as a key to secure systems. As observed, upon performing bit-wise operations on the data read from a page, the number of stable bits is huge even after 50000 operations. An algorithm is designed to tackle this scenario wherein it could run the outer loop for 1,00,000 writes and find the number of stable bits after every 10,000 writes.


 

\begin{thebibliography}{10}

\bibitem{hassan2017internett}
Q.~F. Hassan, S.~A. Madani \emph{et~al.}, \emph{Internet of things: Challenges,
  advances, and applications}.\hskip 1em plus 0.5em minus 0.4em\relax CRC
  Press, 2017.

\bibitem{tehranipoor2018towards}
F.~Tehranipoor, ``Towards implementation of robust and low-cost security
  primitives for resource-constrained iot devices,'' \emph{arXiv preprint
  arXiv:1806.05332}, 2018.

\bibitem{tehranipoor2017exploringg}
F.~Tehranipoor, N.~Karimian, P.~A. Wortman, A.~Haque, J.~Fahrny, and J.~A.
  Chandy, ``Exploring methods of authentication for the internet of things,''
  in \emph{Internet of Things}.\hskip 1em plus 0.5em minus 0.4em\relax Chapman
  and Hall/CRC, 2017, pp. 71--90.

\bibitem{yan2015novelll}
W.~Yan, F.~Tehranipoor, and J.~A. Chandy, ``A novel way to authenticate
  untrusted integrated circuits,'' in \emph{2015 IEEE/ACM International
  Conference on Computer-Aided Design (ICCAD)}.\hskip 1em plus 0.5em minus
  0.4em\relax IEEE, 2015, pp. 132--138.

\bibitem{karimian2019dramnett}
N.~Karimian, F.~Tehranipoor, N.~Anagnostopoulos, and W.~Yan, ``Dramnet:
  Authentication based on physical unique features of dram using deep
  convolutional neural networks,'' \emph{arXiv preprint arXiv:1902.09094},
  2019.

\bibitem{karimian2016evolving}
N.~Karimian, P.~A. Wortman, and F.~Tehranipoor, ``Evolving authentication
  design considerations for the internet of biometric things (iobt),'' in
  \emph{Proceedings of the eleventh IEEE/ACM/IFIP international conference on
  hardware/software codesign and system synthesis}, 2016, pp. 1--10.

\bibitem{karimian2019heartid}
N.~Karimian and F.~Tehranipoor, ``Heartid-based authentication for autonomous
  vehicles using deep learning and random number generators,'' in \emph{Mobile
  Multimedia/Image Processing, Security, and Applications 2019}, vol.
  10993.\hskip 1em plus 0.5em minus 0.4em\relax International Society for
  Optics and Photonics, 2019, p. 109930E.

\bibitem{wortman2020exploring}
P.~A. Wortman, F.~Tehranipoor, and J.~A. Chandy, ``Exploring the coverage of
  existing hardware vulnerabilities in community standards,'' in \emph{Silicon
  Valley Cybersecurity Conference}.\hskip 1em plus 0.5em minus 0.4em\relax
  Springer, 2020, pp. 87--97.

\bibitem{yan2020bittt}
W.~Yan, H.~Zhu, Z.~Yu, F.~Tehranipoor, J.~Chandy, N.~Zhang, and X.~Zhang, ``Bit
  2 rng: Leveraging bad-page initialized table with bit-error insertion for
  true random number generation in commodity flash memory,'' in \emph{2020 IEEE
  International Symposium on Hardware Oriented Security and Trust
  (HOST)}.\hskip 1em plus 0.5em minus 0.4em\relax IEEE, 2020, pp. 91--101.

\bibitem{wortman2018p2mmm}
P.~Wortman, W.~Yan, J.~Chandy, and F.~Tehranipoor, ``P2m-based security model:
  security enhancement using combined puf and prng models for authenticating
  consumer electronic devices,'' \emph{IET Computers \& Digital Techniques},
  vol.~12, no.~6, pp. 289--296, 2018.

\bibitem{tehranipoor2018dvftt}
F.~Tehranipoor, P.~Wortman, N.~Karimian, W.~Yan, and J.~A. Chandy, ``Dvft: A
  lightweight solution for power-supply noise-based trng using dynamic voltage
  feedback tuning system,'' \emph{IEEE Transactions on Very Large Scale
  Integration (VLSI) Systems}, vol.~26, no.~6, pp. 1084--1097, 2018.

\bibitem{eckert2017drnggg}
C.~Eckert, F.~Tehranipoor, and J.~A. Chandy, ``Drng: Dram-based random number
  generation using its startup value behavior,'' in \emph{2017 IEEE 60th
  International Midwest Symposium on Circuits and Systems (MWSCAS)}.\hskip 1em
  plus 0.5em minus 0.4em\relax IEEE, 2017, pp. 1260--1263.

\bibitem{tehranipoor2016robust}
F.~Tehranipoor, W.~Yan, and J.~A. Chandy, ``Robust hardware true random number
  generators using dram remanence effects,'' in \emph{2016 IEEE International
  Symposium on Hardware Oriented Security and Trust (HOST)}.\hskip 1em plus
  0.5em minus 0.4em\relax IEEE, 2016, pp. 79--84.

\bibitem{tehranipoor2017study}
F.~Tehranipoor, N.~Karimian, W.~Yan, and J.~A. Chandy, ``A study of power
  supply variation as a source of random noise,'' in \emph{2017 30th
  International Conference on VLSI Design and 2017 16th International
  Conference on Embedded Systems (VLSID)}.\hskip 1em plus 0.5em minus
  0.4em\relax IEEE, 2017, pp. 155--160.

\bibitem{tehranipoor2017designnn}
F.~Tehranipoor, ``Design and architecture of hardware-based random function
  security primitives,'' 2017.

\bibitem{lyp2021lish}
T.~Lyp, N.~Karimian, and F.~Tehranipoor, ``Lish: A new random number generator
  using ecg noises,'' in \emph{2021 IEEE International Conference on Consumer
  Electronics (ICCE)}.\hskip 1em plus 0.5em minus 0.4em\relax IEEE, 2021, pp.
  1--6.

\bibitem{tehranipoor2018loww}
F.~Tehranipoor, N.~Karimian, P.~A. Wortman, and J.~A. Chandy, ``Low-cost
  authentication paradigm for consumer electronics within the internet of
  wearable fitness tracking applications,'' in \emph{2018 IEEE international
  conference on consumer electronics (ICCE)}.\hskip 1em plus 0.5em minus
  0.4em\relax IEEE, 2018, pp. 1--6.

\bibitem{aguirre2020systematiccc}
A.~Aguirre, M.~Hall, T.~Lim, J.~Trinh, W.~Yan, and F.~Tehranipoor, ``A
  systematic approach for internal entropy boosting in delay-based ro puf on an
  fpga,'' in \emph{2020 IEEE 63rd International Midwest Symposium on Circuits
  and Systems (MWSCAS)}.\hskip 1em plus 0.5em minus 0.4em\relax IEEE, 2020, pp.
  623--626.

\bibitem{anagnostopoulos2018addressingg}
N.~A. Anagnostopoulos, T.~Arul, Y.~Fan, C.~Hatzfeld, F.~Tehranipoor, and
  S.~Katzenbeisser, ``Addressing the effects of temperature variations on
  intrinsic memory-based physical unclonable functions,'' \emph{crypto day
  matters 28}, 2018.

\bibitem{tehranipoor2017investigation}
F.~Tehranipoor, N.~Karimian, W.~Yan, and J.~A. Chandy, ``Investigation of dram
  pufs reliability under device accelerated aging effects,'' in \emph{2017 IEEE
  International Symposium on Circuits and Systems (ISCAS)}.\hskip 1em plus
  0.5em minus 0.4em\relax IEEE, 2017, pp. 1--4.

\bibitem{tehranipoor2016dram}
------, ``Dram-based intrinsic physically unclonable functions for system-level
  security and authentication,'' \emph{IEEE Transactions on Very Large Scale
  Integration (VLSI) Systems}, vol.~25, no.~3, pp. 1085--1097, 2016.

\bibitem{yue2020drammm}
M.~Yue, N.~Karimian, W.~Yan, N.~A. Anagnostopoulos, and F.~Tehranipoor,
  ``Dram-based authentication using deep convolutional neural networks,''
  \emph{IEEE Consumer Electronics Magazine}, 2020.

\bibitem{Holcomb2008}
D.~E. Holcomb, W.~P. Burleson, and K.~Fu, ``Power-up sram state as an
  identifying fingerprint and source of true random numbers,'' \emph{IEEE
  Transactions on Computers}, vol.~58, no.~9, pp. 1198--1210, 2008.

\bibitem{karimian2019generateee}
N.~Karimian and F.~Tehranipoor, ``How to generate robust keys from noisy
  drams?'' in \emph{Proceedings of the 2019 on Great Lakes Symposium on VLSI},
  2019, pp. 465--469.

\bibitem{anagnostopoulos2018securing}
N.~A. Anagnostopoulos, T.~Arul, Y.~Fan, C.~Hatzfeld, J.~Lotichius, R.~Sharma,
  F.~Fernandes, F.~Tehranipoor, and S.~Katzenbeisser, ``Securing iot devices
  using robust dram pufs,'' in \emph{2018 Global Information Infrastructure and
  Networking Symposium (GIIS)}.\hskip 1em plus 0.5em minus 0.4em\relax IEEE,
  2018, pp. 1--5.

\bibitem{anagnostopoulos2018overvieww}
N.~A. Anagnostopoulos, S.~Katzenbeisser, J.~Chandy, and F.~Tehranipoor, ``An
  overview of dram-based security primitives,'' \emph{Cryptography}, vol.~2,
  no.~2, p.~7, 2018.

\bibitem{anagnostopoulos2017insights}
N.~A. Anagnostopoulos, A.~Schaller, Y.~Fan, W.~Xiong, F.~Tehranipoor, T.~Arul,
  S.~Gabmeyer, J.~Szefer, J.~A. Chandy, and S.~Katzenbeisser, ``Insights into
  the potential usage of the initial values of dram arrays of commercial
  off-the-shelf devices for security applications,'' \emph{Proceedings of the
  26th Crypto-Day, Nuremberg, Germany}, pp. 1--2, 2017.

\bibitem{tehranipoor2015dram}
F.~Tehranipoor, N.~Karimian, K.~Xiao, and J.~Chandy, ``Dram based intrinsic
  physical unclonable functions for system level security,'' in
  \emph{Proceedings of the 25th edition on Great Lakes Symposium on VLSI},
  2015, pp. 15--20.

\bibitem{gordon2021flashsh}
H.~Gordon, J.~Edmonds, S.~Ghandali, W.~Yan, N.~Karimian, and F.~Tehranipoor,
  ``Flash-based security primitives: Evolution, challenges and future
  directions,'' \emph{Cryptography}, vol.~5, no.~1, p.~7, 2021.

\bibitem{wortman2017proposing}
P.~A. Wortman, F.~Tehranipoor, N.~Karimian, and J.~A. Chandy, ``Proposing a
  modeling framework for minimizing security vulnerabilities in iot systems in
  the healthcare domain,'' in \emph{2017 IEEE EMBS International Conference on
  Biomedical \& Health Informatics (BHI)}.\hskip 1em plus 0.5em minus
  0.4em\relax IEEE, 2017, pp. 185--188.

\bibitem{yan2017phasee}
W.~Yan, C.~Jin, F.~Tehranipoor, and J.~A. Chandy, ``Phase calibrated ring
  oscillator puf design and implementation on fpgas,'' in \emph{2017 27th
  International Conference on Field Programmable Logic and Applications
  (FPL)}.\hskip 1em plus 0.5em minus 0.4em\relax IEEE, 2017, pp. 1--8.

\bibitem{yan2016pufff}
W.~Yan, F.~Tehranipoor, and J.~A. Chandy, ``Puf-based fuzzy authentication
  without error correcting codes,'' \emph{IEEE Transactions on Computer-Aided
  Design of Integrated Circuits and Systems}, vol.~36, no.~9, pp. 1445--1457,
  2016.

\bibitem{pugazhenthi2019dlaa}
A.~Pugazhenthi, N.~Karimian, and F.~Tehranipoor, ``Dla-puf: deep learning
  attacks on hardware security primitives,'' in \emph{Autonomous Systems:
  Sensors, Processing, and Security for Vehicles and Infrastructure 2019}, vol.
  11009.\hskip 1em plus 0.5em minus 0.4em\relax International Society for
  Optics and Photonics, 2019, p. 110090B.

\bibitem{jia2015extracting}
S.~Jia, L.~Xia, Z.~Wang, J.~Lin, G.~Zhang, and Y.~Ji, ``Extracting robust keys
  from nand flash physical unclonable functions,'' in \emph{International
  Conference on Information Security}.\hskip 1em plus 0.5em minus 0.4em\relax
  Springer, 2015, pp. 437--454.

\bibitem{prabhu2011extracting}
P.~Prabhu, A.~Akel, L.~M. Grupp, S.~Y. Wing-Kei, G.~E. Suh, E.~Kan, and
  S.~Swanson, ``Extracting device fingerprints from flash memory by exploiting
  physical variations,'' in \emph{International Conference on Trust and
  Trustworthy Computing}.\hskip 1em plus 0.5em minus 0.4em\relax Springer,
  2011, pp. 188--201.

\bibitem{cai2017vulnerabilities}
Y.~Cai, S.~Ghose, Y.~Luo, K.~Mai, O.~Mutlu, and E.~F. Haratsch,
  ``Vulnerabilities in mlc nand flash memory programming: Experimental
  analysis, exploits, and mitigation techniques,'' in \emph{2017 IEEE
  International Symposium on High Performance Computer Architecture
  (HPCA)}.\hskip 1em plus 0.5em minus 0.4em\relax IEEE, 2017, pp. 49--60.

\end{thebibliography}

\end{document}